\documentclass[aps,prb,
 amsmath,amssymb,
 twocolumn, groupedaddress,
 showpacs]{revtex4-1}
\usepackage{bm}
\usepackage{graphicx}
\usepackage{gensymb}
\usepackage{amsmath}
\usepackage{amssymb}
\usepackage{color}
\usepackage{ulem}

\begin{document}

\title{Signature of band inversion in the antiferromagnetic phase of axion insulator candidate EuIn$_2$As$_2$}
\author{Takafumi Sato,$^{1,2,3\#}$ Zhiwei Wang,$^{4\#}$ Daichi Takane,$^1$ Seigo Souma,$^{2,3}$ Chaoxi Cui,$^4$ Yongkai Li,$^4$ Kosuke Nakayama,$^{1,5}$ Tappei Kawakami,$^1$ Yuya Kubota,$^1$ Cephise Cacho,$^6$ Timur K. Kim,$^6$ Arian Arab,$^7$ Vladimir N. Strocov,$^7$ Yugui Yao,$^{4*}$ and Takashi Takahashi$^{1,2,3*}$ }

\affiliation{$^1$Department of Physics, Tohoku University, Sendai 980-8578, Japan\\
$^2$Center for Spintronics Research Network, Tohoku University, Sendai 980-8577, Japan\\
$^3$WPI Research Center, Advanced Institute for Materials Research, Tohoku University, Sendai 980-8577, Japan\\
$^4$Key Lab of advanced optoelectronic quantum architecture and measurement (Ministry of Education),\\ Beijing Key Lab of Nanophotonics \& Ultrafine Optoelectronic Systems, and School of Physics, Beijing Institute of Technology, 100081 Beijing, China\\
$^5$Precursory Research for Embryonic Science and Technology (PRESTO), Japan Science and Technology Agency (JST), Tokyo, 102-0076, Japan\\
$^6$Diamond Light Source, Harwell Science and Innovation Campus, Didcot, Oxfordshire OX11 0QX, UK\\
$^7$Swiss Light Source, Paul Scherrer Institut, CH-5232 Villigen, Switzerland
 } 
\collaboration{$^{\#}$These authors contributed equally to this work.}
\collaboration{*Corresponding authors}

\date{\today}

\begin{abstract}
We have performed angle-resolved photoemission spectroscopy on EuIn$_2$As$_2$ which is predicted to be an axion insulator in the antiferromagnetic state. By utilizing soft-x-ray and vacuum-ultraviolet photons, we revealed a three-dimensional hole pocket centered at the $\Gamma$  point of bulk Brillouin zone together with a heavily hole-doped surface state in the paramagnetic phase. Upon entering the antiferromagnetic phase, the band structure exhibits a marked reconstruction characterized by the emergence of a ``M''-shaped bulk band near the Fermi level. The qualitative agreement with first-principles band-structure calculations suggests the occurrence of bulk-band inversion at the $\Gamma$ point in the antiferromagnetic phase. We suggest that EuIn$_2$As$_2$ provides a good opportunity to study the exotic quantum phases associated with possible axion-insulator phase.
\end{abstract}


\maketitle

\section{INTRODUCTION}
Interplay between magnetism and topology is now becoming an exciting topic in condensed-matter physics. Magnetic order that spontaneously breaks the time-reversal symmetry induces various exotic quantum phenomena which are originally absent in time-reversal-invariant $Z_2$ topological insulators (TIs), as represented by the quantum anomalous Hall effect in magnetically doped TIs \cite{YuScience2010, ChangScience2013, CheckelskyNP2014, KouPRL2014, ChangNM2015, QiaoPRB2010}, the unusual electron dynamics in axion TIs \cite{QiPRB2008, LiNP2010}, and the chiral Majorana fermions in superconductor hybrids \cite{QiPRB2010, HeScience2017}. A well-known strategy to realize such fascinating topological phases is introduction of magnetic impurities into TI crystals or fabrication of heterostructures. However, the former inevitably introduces disorders in the crystal and the latter requires complex steps to achieve desired quantum properties. Thus, the realization of intrinsic magnetic TIs where the magnetic order is inherently incorporated into the crystal is highly desired, because it would be a more promising platform to realize ``cleaner'' topological quantum effects.

Amongst various magnetic topological materials, axion insulator is currently attracting a particular attention since it is predicted to host topological magnetoelectric effect associated with a hypothetic quasiparticle called axion. While axion insulator was proposed to be realized in various systems such as heterostructures involving quantum anomalous Hall insulators \cite{MogiNM2017, XiaoPRL2018} and magnetically doped TIs \cite{LiNP2010, YueNP2019}, the experimental investigation of axion insulators with stoichiometric intrinsic magnetic TIs is still limited to a few materials class \cite{LiPRX2019, GuiACS2019, ChenNC2019, GongCPL2019, OtrokovNature2019, HaoPRX2019, ChenArXiv2019, ArXivZhang, ArXivRegmi}. Axion-insulator phase with inherent antiferromagnetic (AFM) order was recently proposed in MnBi$_2$Te$_4$ family \cite{LiPRX2019, ChowdhuryNPJ2019, LiSciAdv2019, ChenNC2019, GongCPL2019, OtrokovNature2019, ZhangPRL2019, LiPRB2019, HaoPRX2019, ChenArXiv2019}; however its axion-insulator nature, e.g. whether or not the surface state (SS) hosts an energy gap associated with the time-reversal-symmetry breaking, is still controversial \cite{LiPRX2019, ChenNC2019, GongCPL2019, OtrokovNature2019, HaoPRX2019, ChenArXiv2019}.

Recently, it was proposed from first-principles band-structure calculations that EuIn$_2$As$_2$ may be a platform of axion insulator exhibiting the coexistence with higher-order topological phases \cite{XuPRL2019}. EuIn$_2$As$_2$ has a layered centrosymmetric crystal structure with space group $P6_3/mmc$ (No.194), and undergoes the paramagnetic (PM) to AFM transition at the N\'eel temperature of $T_N$ = 16 K \cite{GoforthIC2008, SinghAPL2012}; see Figs. 1(a) and 1(b) for the crystal structure and the Brillouin zone (BZ). The AFM state of EuIn$_2$As$_2$ is associated with the Eu$^{2+}$ ions with the magnetic moment of 7.0$\mu_B$, and is characterized by the A-type AFM long-range order where the intra- and inter-layer exchange couplings are ferromagnetic (FM) and AFM, respectively. The AFM state is characterized by the inverted bulk-band structure at the bulk BZ center. As a result of nontrivial topology associated with the parity-based invariant $Z_4$ = 2 \cite{XuPRL2019}, an axion-insulator phase would emerge in the AFM phase. There are two metastable magnetic structures in EuIn$_2$As$_2$, i.e. the magnetic moment aligns parallel to the $a$/$b$ axis (in-plane) or the $c$ axis (out-of-plane), called here AFM${\parallel}b$ or AFM${\parallel}c$ states, respectively. Intriguingly, the AFM${\parallel}b$ state is predicted to host an axion-insulator phase which coexists with a topological-crystalline-insulator (TCI) phase whose high-index crystal surfaces can host hinge states. Moreover, the AFM${\parallel}c$ state also hosts an axion-insulator phase, but it coexists with higher-order TI state with chiral hinge mode \cite{XuPRL2019}. Despite such intriguing theoretical predictions, the electronic state of EuIn$_2$As$_2$ has been scarcely investigated by experiments \cite{ArXivZhang, ArXivRegmi}. It is thus highly desired to clarify the band structure of EuIn$_2$As$_2$.

In this article, we report angle-resolved photoemission spectroscopy (ARPES) study of EuIn$_2$As$_2$ bulk single crystal. By complementary using soft-x-ray (SX) and vacuum-ultraviolet (VUV) photons, we have separately identified three-dimensional (3D) bulk and two-dimensional (2D) surface bands. Temperature-dependent ARPES measurements further revealed an intriguing reconstruction of bulk-band structure associated with the AFM transition. The observed band structure in the AFM phase is consistent with the band-structure calculations, pointing to a possibility that EuIn$_2$As$_2$ in the AFM state hosts the axion-insulator phase.

\begin{figure}
\begin{center}
\includegraphics[width=3.2in]{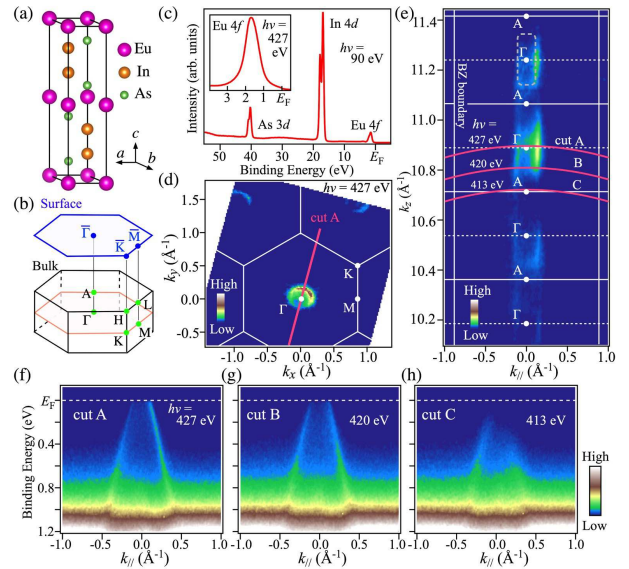}
  \hspace{0.2in}
\caption{(color online). (a) Crystal structure of EuIn$_2$As$_2$. (b) Bulk hexagonal BZ and corresponding surface BZ. (c) EDC in a wide $E_{\rm B}$ region measured with $h\nu$ = 90 eV in the PM phase ($T$ = 30 K). Inset shows the near-$E_{\rm F}$ EDC measured with $h\nu$ = 427 eV. (d) ARPES-intensity mapping at $E_{\rm F}$ as a function of in-plane wave vectors ($k_x$ and $k_y$) at  $k_z$ $\sim$ 0 ($h\nu$ = 427 eV) measured at $T$ = 30 K. (e) ARPES-intensity mapping at $E_{\rm F}$ as a function of  $k_{\parallel}$ and  $k_z$ measured with $h\nu$ = 360-480 eV at $T$ = 30 K, highlighting the bulk holelike FS centered at $\Gamma$ (dashed rectangle). (f)-(h) ARPES intensity as a function of  $k_{\parallel}$ and $E_{\rm B}$ measured along cuts A-C in (e), respectively.
}
\end{center}
\end{figure}

\section{EXPERIMENTAL}
High-quality EuIn$_2$As$_2$ single crystals were synthesized by flux method. ARPES measurements were performed with synchrotron light at SX-ARPES endstation of the ADRESS beamline at the Swiss Light Source, Paul Scherrer Institute, Switzerland \cite{SLS1} and HR-ARPES endstation of I05 beamline in DIAMOND light source, UK. The first-principles calculations were carried out by a projector augmented wave method implemented in Vienna {\it Ab initio} Simulation Package with generalized gradient approximation \cite{VASP, GGA, PBE, MPsample, GGA+U}. We have also calculated the topological SS by means of Green's function method by using the Wannier90 and WannierTools packages \cite{Wannier1, Wannier2}. For details, see Appendix.

\section{RESULTS AND DISCUSSION}
We at first discuss the band structure of EuIn$_2$As$_2$ in the PM phase. Figure 1(c) displays the energy distribution curve (EDC) in a wide energy range measured at photon energy ($h\nu$) of 90 eV at $T$ = 30 K. One can recognize core-level peaks which are assigned to the As-3{\it d}, In-4{\it d}, and Eu-4{\it f} orbitals. The sharp spectral feature and the absence of core-level peaks from other elements confirm the clean sample surface. As shown in the EDC near $E_{\rm F}$ measured at $h\nu$ = 427 eV with enhanced photo-ionization cross section of the Eu 4{\it f} orbital \cite{Lindau}, the Eu 4{\it f} states are located at the binding energy ($E_{\rm B}$) of $\sim$1.7 eV, consistent with the previous band-structure calculations of EuIn$_2$As$_2$ \cite{XuPRL2019} and photoemission data of other Eu-based compounds \cite{GuiACS2019, LiPRX2019}. To elucidate the bulk Fermi-surface (FS) topology, we have mapped out the ARPES intensity as a function of in-plane wave vector with $h\nu$ = 427 eV in the $k_z\sim$0 ($\Gamma$KM) plane of bulk BZ, as shown in Fig. 1(d). The increase of the photoelectron mean free path compared to the VUV energy range reduces, by the Heisenberg relation, the intrinsic uncertainty of the out-of-plane wavevector $k_z$ and thus allows accurate 3D band mapping \cite{Strokov3D}. One can recognize a circular intensity spot centered at $\Gamma$ which follows the periodicity of hexagonal BZ. This is attributed to the hole pocket, as seen from the ARPES intensity in Fig. 1(f) measured along a $k$ cut crossing the $\Gamma$ point [cut A in Fig. 1(d)] which signifies a holelike band crossing $E_{\rm F}$. This hole pocket has a 3D character, as visible from the ARPES-intensity mapping as a function of $k_{\parallel}$ and $k_z$ in Fig. 1(e) [$k_{\parallel}$ is between the $\overline{\Gamma}\overline{M}$ and $\overline{\Gamma}\overline{K}$ cuts of surface BZ; see solid line in Fig. 1(d)] revealed a strong intensity around $\Gamma$ but not around A. Such an intensity variation as a function of  $k_z$ is associated with 3D nature of FS, and is further corroborated with the absence of $E_{\rm F}$-crossing of bands along a $k$ cut passing the A point [cut C in Fig. 1(h)]. We have estimated the size of hole pocket to be 1$\%$ of bulk BZ corresponding to 7$\times$10$^{19}$ cm$^{-3}$ hole carriers, by assuming a 3D cylindrical shape. This indicates the hole-doped nature of bulk EuIn$_2$As$_2$ single crystal as well as a small number of bulk hole carriers, in agreement with the measured positive Hall coefficient and hole concentration of 6.5$\times$10$^{19}$ cm$^{-3}$ estimated from our Hall measurement; see Appendix for details.

\begin{figure}
\includegraphics[width=3.2in]{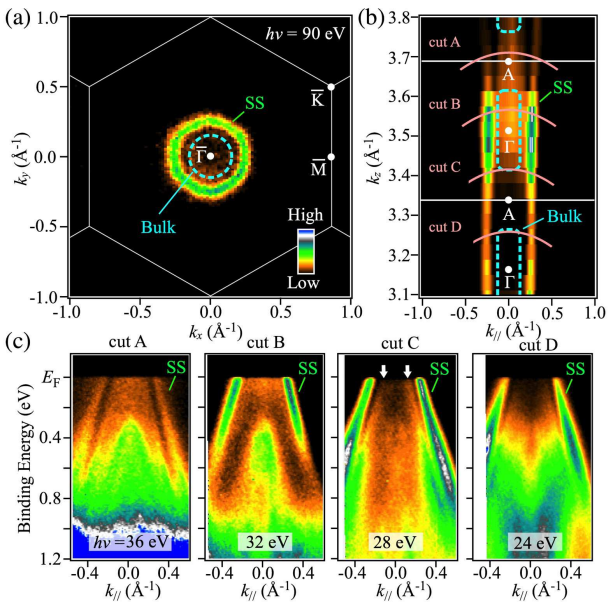}
\hspace{0.2in}
\caption{(color online). (a) ARPES-intensity mapping at $E_{\rm F}$ as a function of in-plane wave vectors measured with VUV photons ($h\nu$ = 90 eV) at $T$ = 35 K. Dashed circle shows the bulk FS at  $k_z$=0 plane determined with SX photons (see Fig. 1d). (b) ARPES-intensity mapping at $E_{\rm F}$ as a function of  $k_{\parallel}$ and  $k_z$ measured with $h\nu$ = 20-40 eV at $T$ = 35 K. Dashed curve shows the bulk FS determined with SX photons; see Fig. 1(e). (c) ARPES intensity as a function of  $k_{\parallel}$ and $E_{\rm B}$ measured along cuts A-D in (b).
}
\end{figure}

To clarify the existence of SS, we performed ARPES measurements with surface-sensitive VUV photons. Figure 2(a) displays the ARPES-intensity mapping at $E_{\rm F}$ as a function of in-plane wave vector in the PM phase ($T$ = 35 K) measured at $h\nu$ = 90 eV. One finds a circular pocket centered at $\bar{\Gamma}$, similarly to the case of SX photons [Fig. 1(d)]. However, taking a careful look one can find that this FS is much bigger than that of bulk FS at  $k_z$ $\sim$ 0 (dashed circle), suggestive of its surface origin. In fact, the ARPES intensity shown in Fig. 2(b) as a function of  $k_{\parallel}$ and  $k_z$ displays a fairly straight FS segment unlike the case of 3D bulk FS (dashed curves), supportive of its 2D nature. As shown in Fig. 2(c), this SS forms an outermost linearly dispersive holelike band with the Fermi vector ($k_{\rm F}$) robust against $h\nu$ variation. On the other hand, one can recognize a much weaker intensity inside this holelike SS. This feature exhibits a finite $h\nu$ variation and its  $k_{\rm F}$ point (white arrows in cut C) is consistent with that of bulk observed with SX photons. The observed large difference in the  $k_{\rm F}$ points between the bulk and surface states suggests extra hole doping to the surface, likely caused by the abrupt termination of crystal at the surface \cite{GuiACS2019, LiPRX2019}. 

\begin{figure*}
\includegraphics[width=6.3in]{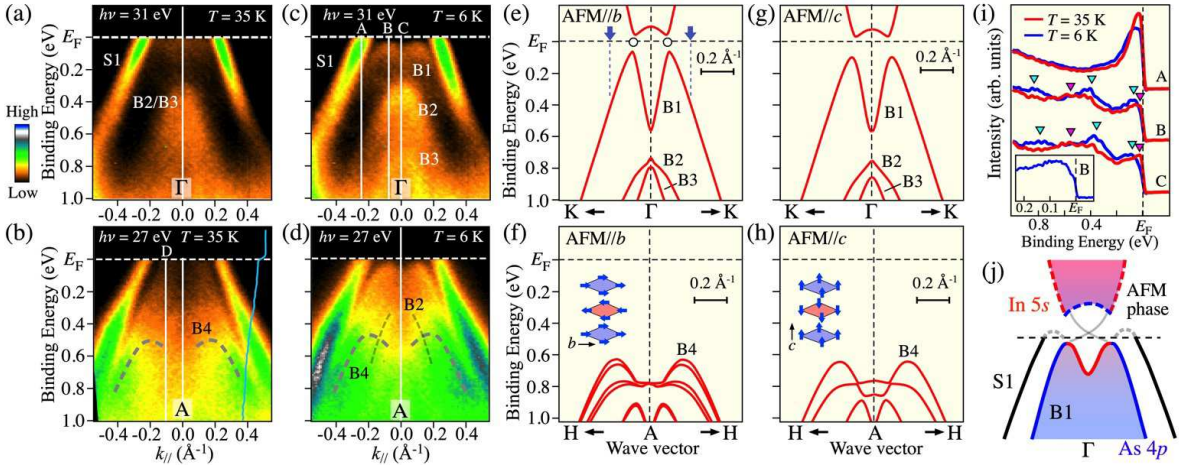}
\hspace{0.2in}
\caption{(color online). (a),(b) ARPES intensities as a function of  $k_{\parallel}$ and $E_{\rm B}$ at  $k_z$ $\sim$ 0 ($h\nu$ = 31 eV) and  $k_z$ $\sim$ $\pi$ ($h\nu$ = 27 eV), respectively, measured at $T$ = 35 K. EDC at point D is also plotted by light blue curve in (b) to show the existence of a Fermi edge supporting the $E_{\rm F}$-crossing of the weak holelike feature. (c),(d) Same as (a) and (b), respectively, but measured at $T$ = 6 K. (e),(f) Band structures obtained from first-principles band structure calculations for the AFM${\parallel}b$ state [see inset to (f)], for  $k_z$ = 0 and $\pi$, respectively. (g),(h) Same as (e) and (f) but for the AFM${\parallel}c$ state. (i) EDCs at $T$ = 6 and 35 K measured at three representative $k$ points (A-C) indicated by white solid lines in (c). EDC at $T$ = 6 K magnified around $E_{\rm F}$ at point B is shown in the inset. (j) Schematic band picture in the AFM phase for bulk In 5{\it s}, As 4{\it p} states and SS deduced from the present ARPES experiment.
}
\end{figure*}

Next, we address a key question regarding how the observed band structure is influenced by the AFM transition. Figures 3(a) and 3(b) show the ARPES intensity at $T$ = 35 K (in the PM phase) measured along $k$ cuts crossing the $\Gamma$ point at  $k_z$ $\sim$ 0 and the A point at  $k_z$ $\sim$ $\pi$, respectively, which again confirms an intense holelike SS (called here S1 band) and a broad bulk band inside the SS. When the crystal was cooled down to 6 K, well below the AFM transition temperature ($T_N$ = 16 K), we observed a marked change in the electronic states. As shown by a comparison of ARPES intensity at  $k_z$ $\sim$ 0 between the PM and AFM phases in Figs. 3(a) and 3(c), a weak broad feature near $E_{\rm F}$ around $\Gamma$ transforms into a more intense ``M''-shaped band (called B1) within 0.1 eV of $E_{\rm F}$, whereas the S1 band shows no discernible change across $T_N$. Moreover, the M-shaped band does not cross $E_{\rm F}$ and a finite energy gap opens at $E_{\rm F}$ [note that such a gap opening can be also visible from a peak shift of EDC at point B; see Fig. 3(i)]. One can also observe another drastic change in the second and third holelike bands called B2 and B3, which degenerate at $\Gamma$ in the PM phase, but apparently split in the AFM phase. At  $k_z$ = $\pi$, a similar M-shaped band is resolved in the AFM phase [Fig. 3(d)]. This band becomes a broad holelike feature crossing $E_{\rm F}$ in the PM phase while its intensity is weak [Fig. 3(b)]. Such temperature-dependent change is well visualized in the EDCs at representative $k$ points [A-C in Fig. 3(c)] where a number of peaks and its location are apparently different at point B (top of the B1 band) and point C (the $\Gamma$ point) in Fig. 3(i). These results demonstrate a remarkable reconstruction of bulk electronic states associated with the AFM transition. It is noted that the energy dispersion of S1 band is insensitive to the AFM transition, as visible from no discernible shift in the peak energy at point A (the  $k_{\rm F}$ point of S1 band) in Fig. 3(i). This suggests a weak interaction between the SS and the localized magnetic moment.

 To understand in more detail the observed peculiar electronic states in the AFM phase, we have performed bulk-band-structure calculations for the AFM phase. Since there are two metastable magnetic structures in the AF phase (AFM${\parallel}b$ and AFM${\parallel}c$) and their total energies are very similar to each other \cite{XuPRL2019}, we have calculated the band structure for both cases and display the results in Figs. 3(e)-3(h) (note that the calculation for the PM phase is technically difficult because the standard density-functional-theory (DFT) codes that deal with the state at $T$ = 0 K have a difficulty in simulating the random spin state at $T >$ 0 K in the PM phase). One can notice that the global band structure is very similar between the two phases. For instance, one commonly finds an inverted band structure where the electronlike In 5{\it s} conduction band and the As 4{\it p} holelike valence band at  $k_z$ = 0 ($\Gamma$K cut) largely overlap each other around $\Gamma$, showing an energy gap of 0.1 eV at the intersection due to the spin-orbit coupling. As a result, the topmost occupied electronic states show a M-shaped band dispersion. Such insensitivity of the low-lying bands to the magnetic structure is associated with the fact that the Eu 4{\it f} state is located far away from $E_{\rm F}$. As shown in Figs. 3(f) and 3(h), the As 4{\it p} bands move away from $E_{\rm F}$ at  $k_z$ = $\pi$. Besides such similarity of calculated band structure between the two phases, one finds some differences between them, such as a small band splitting along the AH cut for the AFM${\parallel}b$  phase; this is attributed to the difference in the magnetic space group ($Cmcm$ vs $P6_{3'}/m'm'c$) \cite{XuPRL2019}.

Now we discuss the characteristics of observed bands by comparing the band structure between experiments and calculations. Obviously, the S1 band observed by ARPES has no counterpart in the bulk-band calculation because the  $k_{\rm F}$ point of the S1 band [arrows in Fig. 3(e); 0.20 \AA$^{-1}$ from the $\Gamma$ point] is far from the intersection of the bulk In 5{\it s} and As 5{\it p} bands [open circles; 0.12 \AA$^{-1}$ from the $\Gamma$ point]. One may argue that such discrepancy is a consequence of heavy hole doping into the bulk crystal which results in a chemical-potential ($\mu$) shift of $\sim$0.4 eV (see vertical dashed lines). In this case, the S1 band may coincide with the bulk band. However, such a large chemical-potential shift is unreasonable when we take into account the tiny hole carriers in the bulk crystal estimated from our Hall measurement (6.5$\times$10$^{19}$ cm$^{-3}$). The reasonable agreement in the $k$ position of M-shaped band-top between the experiment and calculations without discernible $\mu$-shift, together with the apparent inconsistency in the size of bulk and surface-derived FSs in Figs. 1(d) and 2(a), strongly suggests that the observed M-shaped band is of bulk origin and the bulk $\mu$-shift is not so large in contrast to the highly hole-doped nature of the SS.
     
Based on above band assignment, one can further observe some quantitative differences in the location of bands between the experiment and calculations. For instance, the bottom of electronlike In 5{\it s} states at $\Gamma$ is much deeper in the calculation; i.e, the energy overlap between the In 5{\it s} and As 4{\it p} states is significantly overestimated in the calculation. However, it should be emphasized that, despite such a quantitative difference, the experimental band structure still maintains the inverted character (i.e., M-shaped feature), which is a prerequisite for the realization of predicted topological phases in the AFM state \cite{XuPRL2019}. One can see from a direct comparison of Figs. 3(c) and 3(e, g) that the experimental B2 band is pushed upward by $\sim$0.4 eV compared to the calculated counterpart topped at $E_{\rm B}$ $\sim$ 0.8 eV. Such a shift would be correlated with the shallower In 5{\it s} electronlike feature in the experiment.

As shown by a comparison of the electronic states at  $k_z$ = $\pi$ in Figs. 3(d) and 3(f, h), the ARPES result and the calculated band structure share a common feature, i.e. a characteristic non-parabolic band dispersion called here B4 [see dashed curve in Figs. 3(b) and 3(d)]. Since such a non-parabolic band is absent in the experiment at  $k_z$ $\sim$ 0 as seen in Fig. 3(c), it is inferred that the VUV-ARPES data certainly reflect the intrinsic  $k_z$ variation in the bulk-band structure. On the other hand, one can commonly recognize the M-shaped band near $E_{\rm F}$ in both Figs. 3(c) and 3(d). A weak signature of the holelike band that resembles the dispersion of the B2 band can be also seen in Fig. 3(d) (thin dashed curve). This is unexpected from the calculated band structure at  $k_z$ = $\pi$ that shows a large gap of $\sim$0.6 eV below $E_{\rm F}$, suggesting that the ARPES intensity is influenced by both the  $k_z$-broadened and  $k_z$-selective spectral weights. Such a large  $k_z$-broadening effect is likely due to the short escape depth of photoelectrons excited by VUV photons together with the small BZ size along  $k_z$ (0.35 \AA$^{-1}$), as also corroborated by observation of a M-shaped band in a wide VUV-photon-energy range ($h\nu$ = 20-40 eV). It is also inferred that the ARPES intensity suffers strong matrix-element effect, judging from the observation that the holelike band which is obvious in the SX data in Fig. 1(f) is not clearly visible in the VUV data in Fig. 3(a). Appearance of the M-shaped band in the AF phase despite the strong intensity suppression in the PM phase suggests an interesting possibility that the matrix-element effect is different between the PM and AF phases in the region where the strong band reconstruction takes place. It is also remarked here that the linewidth of ARPES spectrum is not sharp enough to distinguish a possible difference in the band structure expected from the calculations for AFM${\parallel}b$ and AFM${\parallel}c$ phases. It is necessary to pin down the actual magnetic structure from other experiments such as neutron diffraction.

Now we discuss implications of the present results in relation to the predicted topological phases. As illustrated in the band diagram in Fig. 3(j), the ARPES data combined with the band calculations suggest the occurrence of bulk-band inversion at $\Gamma$  in the AFM phase. The Fermi level is located close to, but slightly above the valence-band maximum due to the hole doping. While the observed S1 band may merge into the bulk conduction or valence bands, their actual connectivity is unclear due to the hole-doped nature. Nevertheless, it is inferred that the S1 band is a trivial SS originating from the As-terminated surface and would likely be connected to the valence band [see Fig. 3(j)], since a similar SS was predicted for the P-terminated surface of a sister compound EuSn$_2$P$_2$ and confirmed by ARPES \cite{GuiACS2019}.  We speculate that, besides the S1 band, another topological Dirac-cone SS that is connected to the B1 band may emerge in mainly above-$E_{\rm F}$ region in EuIn$_2$As$_2$ (schematically shown by gray curves) \cite{XuPRL2019} since the band inversion must be accompanied by the Dirac-cone SS within the spin-orbit gap. Existence of such topological SS is also supported by our first-principles band calculations by means of Green's function method; see Appendix for details. A careful look at the EDC at $T$ = 6 K at point B shown in the inset to Fig. 3(j) suggests that there exists a Fermi edge cut-off besides a hump slightly away from $E_{\rm F}$ originating from the M-shaped band. This suggests that the observed gap at point B is not a full gap but an imperfect gap. This may be consistent with the appearance of a weak topological SS crossing $E_{\rm F}$ although this conjecture should be verified by visualizing the full energy dispersion by experiments such as two-photon ARPES.

According to the theory \cite{XuPRL2019}, the predicted axion-insulator phase coexisting with the exotic TCI (for AFM${\parallel}b$) or higher-order TI (for AFM${\parallel}c$) requires an inverted band structure at $\Gamma$ as well as a gapped Dirac-cone band on some surfaces. Thus, from the present ARPES result, it can be said that EuIn$_2$As$_2$ in the AFM phase is either a TI or an axion insulator, depending upon whether or not a finite magnetic gap opens at the Dirac point in the AFM phase. Clarification of the actual magnetic structure and observation of the Dirac-cone SS modulated by the antiferromagnetism are the next important step to further investigate the exotic physical properties associated with the topological characteristics of EuIn$_2$As$_2$.
    
\section{SUMMARY}
In conclusion, the present ARPES study of EuIn$_2$As$_2$ has revealed a band reconstruction associated with the AFM transition together with a signature of bulk-band inversion in the AFM phase, consistent with the band calculation that supports the axion-insulator phase in EuIn$_2$As$_2$. The present study lays a foundation for understanding the interplay between antiferromagnetic order and non-trivial topology.

\begin{acknowledgments}
We thank T. Saito, K. Horiba, M. Kitamura, and H. Kumigashira for their assistance in the ARPES experiments. This work was supported by Grant-in-Aid for Scientific Research on Innovative Areas ``Topological Materials Science'' (JSPS KAKENHI Grant Number JP15H05853 and No. JP15K21717), JST-CREST (No. JPMJCR18T1), JST-PRESTO (No. JPMJPR18L7), and Grant-in-Aid for Scientific Research (JSPS KAKENHI Grant Numbers JP17H01139, JP18H04472, JP18H01160, JP18J20058, and JP17H04847), KEK-PF (Proposal number 2018S2-001), SLS (Proposal number 20190829), DIAMOND light source (proposal number SI23799). The work at Beijing was supported by the Natural Science Foundation of China (NSFC Grant No. 11734003), the National Key R$\&$D Program of China (Grant No.2016YFA0300600), the Strategic Priority Research Program of Chinese Academy of Sciences (Grant No. XDB30000000), Z.W. acknowledges the support from Beijing Institute of Technology Research Fund Program for Young Scholars. A.A. acknowledges Swiss National Science Foundation (Grant No. 200020B\_188709).
\end{acknowledgments}

\appendix
\section{SAMPLE FABRICATION, ARPES EXPERIMENTS, AND BAND CALCULATIONS}
High-quality EuIn$_2$As$_2$ single crystals were synthesized by flux method. High-purity Eu (lump), In (shot) and As (lump) were loaded into an Al$_2$O$_3$ crucible with the atomic ratio of Eu:In:As = 1:12:3, and sealed into a quartz tube in a vacuum of 2 $\times$ 10$^{-6}$ Torr. The tube was heated up to 1373 K in 20 h and held for 10 h at this temperature, then slowly cooled down to 973 K at a rate of 2 K/h, at which the flux was removed by centrifuge. Shiny crystals with typical size of 2 $\times$ 2 $\times$ 0.2 mm$^3$ were obtained. 

SX and VUV ARPES measurements were performed with energy-tunable synchrotron light at SX-ARPES end-station of the ADRESS beamline at the Swiss Light Source (SLS), Paul Scherrer Institute, Switzerland \cite{SLS1} and HR-ARPES end-station of I05 beamline in the DIAMOND light source, UK, respectively. SX- and VUV-ARPES measurements were performed with 320-480 and 20-120 eV photons with $p$/circular and $p$-polarizations, respectively. The energy resolutions for SX- and VUV-ARPES measurements were 40-60 and 6-25 meV, respectively. Samples were cleaved {\it in situ} along the (001) crystal plane in an ultrahigh vacuum of 1 $\times$ 10$^{-6}$ Torr.
The first-principles calculations were carried out by a projector augmented wave method implemented in Vienna {\it Ab initio} Simulation Package (VASP) \cite{VASP} with generalized gradient approximation (GGA) \cite{GGA} and the Perdew-Burke-Ernzerhof (PBE) \cite{PBE} type exchange-correlation potential. Fully optimized lattice constants of $a$ = $b$ = 4.2951 \AA, and $c$ = 17.9583 \AA, were used for the further calculations. The spin-orbit coupling was included self-consistently. A uniform grid of 13 $\times$ 13 $\times$ 5 was used for sampling the Brillouin zone \cite{MPsample}. The energy cutoff was set as 400 eV. To better process the on-site Coulomb interactions, the Hubbard $U$ parameter was set as 5 eV for Eu 4$f$ electrons in the GGA+$U$ calculations \cite{GGA+U}.

\section{HALL MEASUREMENT}

Figure 4 shows the magnetic field dependence of $\rho_{yx}$ at $T$ = 2 K (black) and 300 K (red). One can see that the $\rho_{yx}$ shows a linear dependence on magnetic field regardless of temperature. The positive slope indicates that the dominant carrier is hole in this system. The hole density $p$ at $T$ = 2 K was estimated to be 6.5 $\times$ 10$^{19}$ cm$^{-3}$ from the relationship $p$ = 1/$eR_H$, where $R_H$ is the Hall coefficient and $e$ is the elementary charge.

\begin{figure}
 \includegraphics[width=3 in]{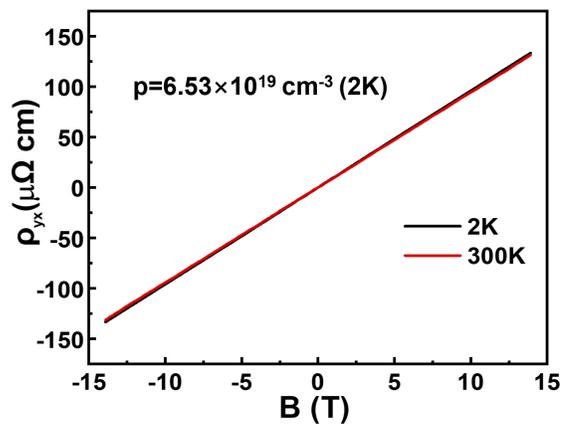}
\caption{(color online). Magnetic field dependence of $\rho_{yx}$ at $T$ = 2 K (black) and 300 K (red).}
\end{figure}

\begin{figure}
 \includegraphics[width=3.2in]{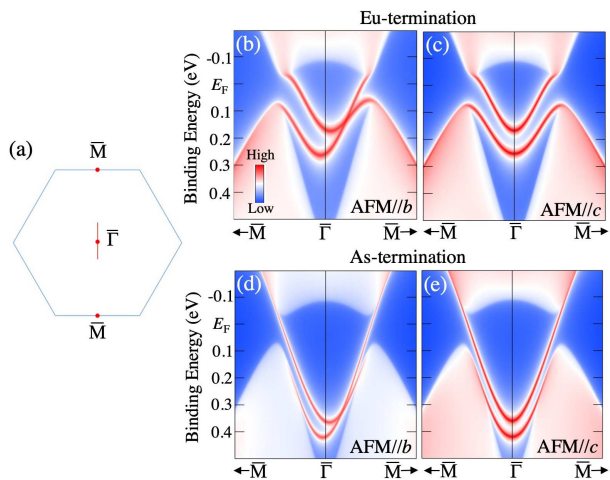}
\caption{(color online). (a) Surface hexagonal Brillouin zone of EuIn$_2$As$_2$ and a $k$ cut (red line) where the calculation was performed. (b), (c) Calculated ARPES intensity profiles along the $\overline{\Gamma}\overline{M}$ cut for the Eu-terminated surface in the AFM${\parallel}b$ and AFM${\parallel}c$ phases, respectively. (d), (e) Same as (b) and (c), respectively, but for the As-terminated surface.}
\end{figure}

\section{BAND-STRUCTURE CALCUATIONS FOR THE TOPOLOGICAL SURFACE STATE}
To numerically simulate the topological SS, we have carried out first-principles band-structure calculations by means of Green's function method by using Wannier90 package \cite{Wannier1} to generate the maximally localized Wannier functions for $d$ and $f$ orbitals of Eu, $s$ and $p$ orbitals of In, and $p$ orbitals of As. By using WannierTools package \cite{Wannier2}, we have calculated the energy dispersion of the topological SS on the (001) surface for two types of AFM order (AFM${\parallel}b$ and AFM${\parallel}c$). As shown in Figs. 5(b) and 5(c), the calculated ARPES intensity profile around the $\overline{\Gamma}$ point along the $\overline{\Gamma}\overline{M}$ cut for the Eu-terminated surface [red line in Fig. 5(a)] signifies a broad intensity distribution from the M-shaped bulk valence band below $E_{\rm F}$ and the W-shaped bulk conduction band above $E_{\rm F}$ (white and light red region), consistent with the bulk-band calculation shown in Figs. 3(e)-(h) of the main text. One can recognize in Fig. 5(b) a couple of sharp dispersive bands (red curves) in the projection gap of the bulk bands (blue region) that traverse the gap between bulk valence- and conduction-bands. These bands are attributed to the topological Dirac-cone SS associated with the bulk-band inversion. For AFM${\parallel}b$, a largely deformed gapless Dirac-cone band is found to be shifted along the in-plane wave vector [Fig. 5(b)], while, for AFM${\parallel}c$, a finite energy gap opens at the Dirac point. This is qualitatively similar to the previous report \cite{XuPRL2019}, and theoretically supports the axion-insulator phase in the AFM${\parallel}c$ state of EuIn$_2$As$_2$. To examine the influence of surface termination on the calculated band structure, we have also carried out the calculation for the As-terminated surface, and show the result in Figs. 5(d) and 5(e). One can recognize that the shifted and gapped Dirac cones for the AFM${\parallel}b$ and AFM${\parallel}c$ phases, respectively, are also seen in this calculation, whereas the Dirac point for both AFM${\parallel}b$ and AFM${\parallel}c$ phases sinks well below $E_{\rm F}$ relative to that of the Eu-terminated surface [Fig. 5(b) and 5(c)]. This suggests that the energy position of topological SS relative to that of the bulk band is highly surface-condition dependent.

\bibliographystyle{prsty}

\end{document}